\documentclass[11pt,preprintnumbers,aps,amssymb,nofootinbib,amsmath,superscriptaddress,notitlepage,prd]
{revtex4-2}
\usepackage{epsfig,epsf}
\usepackage{comment}
\usepackage{bm} 
\usepackage{color} 
\usepackage{slashed} 
\usepackage{relsize}	
\usepackage{soul} 
\usepackage{hyperref}
\usepackage{tensor} 
\usepackage{yfonts} 
\newcommand{\beq}{\begin{equation}}
\newcommand{\beql}[1]{\begin{equation}\label{#1}}
\newcommand{\eeq}{\end{equation}}
\def\bal#1\gal{\begin{align}#1\end{align}}
\newcommand{\ball}[1]{\bal\label{#1}}
\newcommand{\eq}[1]{(\ref{#1})}
\newcommand{\fig}[1]{Fig.~\ref{#1}}

\renewcommand{\b}[1]{{\bm #1}} 

\usepackage{feynmp-auto}

\begin{document}

\title{Electromagnetic radiation at extreme angular velocity}

\author{Matteo Buzzegoli}
\author{Kirill Tuchin}

\affiliation{
Department of Physics and Astronomy, Iowa State University, Ames, Iowa, 50011, USA}

\date{\today}

\begin{abstract}

We consider a system rotating at  extremely high angular velocity, so that its matter is found mostly at the light-cylinder. We posit that it can be described by quantum fields confined to the two-dimensional  cylindrical surface rotating about its symmetry axis. We apply this model to study the electromagnetic radiation. In particular, we compute the photon spectrum emitted by the quark-gluon plasma.

\end{abstract}

\maketitle
\section{Introduction}\label{sec:intro}

The interest to the rapidly rotating systems has been recently rekindled thanks to the experimental observation of highly vortical quark-gluon plasma produced in the relativistic heavy-ion collisions \cite{STAR:2017ckg, Adam:2018ivw, Adam:2019srw, ALICE:2019aid, STAR:2020xbm, STAR:2021beb}.
Previous studies discussed the thermodynamics and the hydrodynamics of rotating systems based on the statistical approach \cite{zubarev,VilenkinQFT,Hayata:2015lga,Ambrus:2015lfr,Buzzegoli:2017cqy,Becattini:2019dxo,Montenegro:2020paq}, developed a quantum kinetic theory with spin degrees of freedom \cite{Weickgenannt:2019dks,Gao:2019znl,Hattori:2019ahi,Bhadury:2020cop,Liu:2020flb,Weickgenannt:2020aaf,Shi:2020htn,Weickgenannt:2022zxs,Hidaka:2017auj}, and made predictions for the
spin polarization measured in heavy-ion collisions \cite{Karpenko:2016jyx,Xie:2017upb,Wu:2019eyi,Ivanov:2020udj,Fu:2020oxj,Li:2017slc,Wei:2018zfb,Ivanov:2019ern,Guo:2021udq,Alzhrani:2022dpi}, see for instance the reviews \cite{Florkowski:2018fap,LNPRotation:2021,Becattini:2020ngo}.

In \cite{Buzzegoli:2022dhw,Buzzegoli:2023vne} we initiated a study of the electromagnetic radiation emitted by rapidly rotating systems. The advantage of the electromagnetic radiation is that it is only weakly affected by the plasma evolution. The idea is to observe the impact of rotation on the quantum fields. 

The study in \cite{Buzzegoli:2022dhw,Buzzegoli:2023vne} discussed ``relatively slowly" rotating systems in magnetic field. Namely, we assumed that the magnetic length $1/\sqrt{eB}$ is  much shorter than the inverse angular velocity $\Omega$\footnote{We are using natural units where $c=\hbar=k_B=1$.}. Such  rotation  is slow. On the other hand, the absolute value of the angular velocity satisfying this condition can be enormous, hence the qualifying adverb ``relatively". Generally, we can say that a system is relatively slowly rotating if its transverse size $a$ is much smaller than $1/\Omega$.

Model simulations show that the vorticity of the quark-gluon plasma can be as high as its inverse transverse size $a$. This upsets the slow rotation assumption. 
A system rotating with the angular velocity $\Omega$ is causally connected only within the lightcone cylinder of radius $R=1/\Omega$. When $R<a$ only a part of the rotating plasma is causally connected. This is a genuine fast rotation. Setting the proper boundary conditions on the quantum fields at the causal boundary becomes an essential procedure. In \cite{Buzzegoli:2023eeo}, in the spirit of the MIT bag model, we required that the radial current vanishes on the boundary. However, there may be other possible boundary conditions. 

In the regime $1/\Omega< \ell$, where $\ell$ is the mean-free-path, the rotation is so fast that it overwhelms all inter-particle forces and pushes the matter towards the light-cylinder wall. Such a medium  will break down to a set of rotating cylindrical regions of radii $R\ll a$. Within each cylindrical region the matter will be concentrated mostly at the boundary at $R$ due to the centrifugal force.  It seems reasonable therefore, that the dynamics of such extremely rapidly rotating system  can be described by the quantum fields confined to the cylindrical surface of radius $R$. We will not be concerned with the statistical properties of the matter within the cylindrical region, since in view of $\ell>R$, it is simply an ideal rotating gas. Rather we are interested in the electromagnetic radiation it emits.  Since the precise nature of the particles that make up the rotating cylinder is not very important --- only the fact that they are found at the boundary is --- we employ the scalar QED  for calculations. It is reasonable to expect that the qualitative features of our results should be fairly model-independent. 

The three rotation regimes of a system characterized by the radial size $a$, the light-cylinder radius $R=1/\Omega$, the mean-free-path $\ell$ are
\begin{itemize}
\item \emph{Slow rotation} $\ell\ll a\ll R$. This approximation is used in \cite{Buzzegoli:2022dhw,Buzzegoli:2023vne}.

\item \emph{Fast rotation} $\ell\ll R\sim a$ \cite{Buzzegoli:2022omv,Buzzegoli:2023eeo}.

\item \emph{Extremely fast rotation} $R\ll \ell\ll a$. This is the scenario we consider in the paper. 
\end{itemize}

In summary, we consider model in which the charged scalar particles can freely move on a thin cylindrical sheet of radius $R=1/\Omega$ rotating with angular velocity $\Omega$. In the following sections we will compute the electromagnetic radiation by a single particle and by a system of particles in thermal equilibrium. We believe that this model describes the universal properties of extremely rapidly rotating systems.

\section{Radiation by scalar particle on cylindrical sheet}

The wave function of a scalar particle of mass $M$ embedded into a cylindrical surface rotating with the angular velocity $\Omega$ about its symmetry axis $z$ is
\ball{a1}
\psi(t,\phi,z)= \frac{1}{\sqrt{2\pi L}}\frac{1}{\sqrt{2E}}e^{-iEt+ip_zz+im \phi}\,,
\gal
where energy spectrum is
\ball{a2}
E=\sqrt{p_z^2+\frac{m^2}{R^2}+M^2}+m\Omega\,.
\gal
The magnetic quantum number $m$ is an integer, while the longitudinal momentum $p_z$ is continuous assuming that  the cylinder height $L$ is very large. We note that $E> 0$ for any $m$. 

\begin{figure}[ht]
    \centering
    \includegraphics[height=4.5cm]{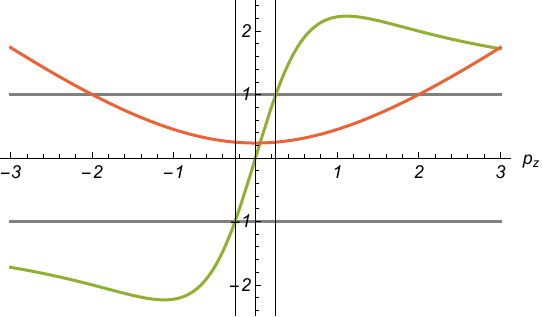}
    \caption{(Color online) $m\Omega =-2$, $M=1$. Green line: $v(p_z)$, red line: $E(p_z)$. The allowed values: $-\frac{1}{4}\le p_z\le \frac{1}{4}$ corresponding to the box in the center.   }
    \label{fig:f1}
\end{figure}

The velocity of the quasi-classical motion along the $z$-direction is  $v=p_z/E$. Clearly, only the states with $|v|\le 1$ are causally connected. Inspection of \eq{a2} reveals that when  $m\ge 0$, then $E>|p_z|$ for any value of $p_z$. In contrast, when $m<0$ this condition is satisfied only when 
\ball{a15}
|p_z|\le \frac{M^2}{2|m|\Omega}\,, \quad \text{if} \quad m<0\,.
\gal
\fig{fig:f1} shows an example of the dispersion relation \eq{a2} with $m<0$. Vertical lines indicated the allowed range of  $p_z$'s. 

We are interested to compute the electromagnetic radiation by this particle. The $S_{fi}$-matrix element reads
\ball{a3}
S_{fi}= -ie\int dt \int d\phi\int dz\, \b j_{fi}(t,\phi,z)\cdot \b A^*(t,\phi,z,R)\,,
\gal
where the photon wave function in the Coulomb gauge is 
\ball{a4}
\b A(t,\phi,z,R)= \frac{1}{\sqrt{2\omega V}}\b \epsilon_\lambda e^{i\b k\cdot \b r-i\omega t }\,,\quad \b\epsilon_\lambda\cdot \b k =0\,,
\gal
and $V$ is the system volume. The transition current is
\ball{a5}
\b j_{fi}= i(\psi \b\nabla \psi'^*-\psi'^*\b \nabla \psi)\,.
\gal
In our notation: $\psi_i=\psi$, $\psi_f= \psi'$.

Substituting \eq{a1} and \eq{a4}  into \eq{a3} we arrive at 
\ball{b1}
S_{fi}=& \frac{-ie (2\pi)^2}{2\pi L\sqrt{2\omega V}2\sqrt{EE'}}\delta(E-E'-\omega)\delta(p_z-p_z'-k_z)\nonumber\\
&\times
\b \epsilon^*_\lambda\cdot \left[ \frac{1}{R}\b e_\phi(m+m')+\b e_z(p_z+p_z')\right]2\pi i J_{m-m'}(k_\bot R)\,.
\gal
Two convenient photon polarizations (following \cite{Sokolov:1986nk}): 
\ball{b2}
\b\epsilon_1= -\b e_\phi\,,\quad 
\b\epsilon_2= -\sin\theta \b e_z+\cos\theta \b e_\bot\,.
\gal
where $\theta$ is the polar angle defined with respect to the $z$-axis, e.g.\ $k_z= \omega \cos\theta$. The photon transverse momentum is then $k_\bot =\omega \sin\theta$.

The photon emission rate can be computed as 
\bal
\dot w_{fi}&=\sum_\lambda\sum_{m'}\frac{|S_{fi}|^2}{T}\frac{dk_z L}{2\pi}\frac{dk_\bot k_\bot \pi R^2}{2\pi}\frac{dp_z L}{2\pi}\label{b4}
\gal 
while the radiation intensity is given by $W= \dot w_{fi} \omega$:
\bal
W&=\sum_{m'=-\infty}^{m-1}\frac{e^2}{16\pi  E E'}\delta(E-E'-\omega)\left[ \frac{(m+m')^2}{R^2}+\sin^2\theta(p_z+p_z')^2\right]J_{m-m'}^2(k_\bot R)dk_z dk_\bot k_\bot\,. \label{b5}
\gal
The sum over the negative $m'$ in \eq{b5} is constrained by \eq{a15}.
One can expresses  $dk_zdk_\bot k_\bot=\omega^2d\omega d\sin\theta = \omega^2d\omega do/2\pi$, where $do$ is the element of the solid angle in the direction of the emitted photon. 
In the non-relativistic limit, the leading term in the multipole expansion of the intensity is the magnetic dipole one because the electric dipole moment vanishes while the magnetic moment $\b\mu= \frac{1}{2}e\b r\times \b v$ is finite.

The delta-function in \eq{b5} can be re-written as 
\ball{b6}
\delta(E-E'-\omega)= \frac{\delta(\omega-\omega_0)(E'-m'\Omega)}{E-m'\Omega-\omega \sin^2\theta-p_z\cos\theta}\,.
\gal
where the characteristic frequency is
\ball{b7}
\omega_0&=\frac{1}{\sin^2\theta}\bigg\{ 
E-m'\Omega-p_z\cos\theta
\nonumber\\
&
-\sqrt{(E-m'\Omega-p_z\cos\theta)^2-
\sin^2\theta\left[ (E-m'\Omega)^2-p_z^2-m'^2/R^2-M^2\right]}\bigg\}\,.
\gal
Taking the integral over $\omega$ one is left with the angular spectrum $dW/do$.  Alternatively, we can cast the delta-function in the form  
\ball{b8}
\delta(E-E'-\omega)= \sum_{\pm}\frac{\delta(\cos\theta-\cos\theta_\pm)(E'-m'\Omega)}{\omega|\omega\cos\theta-p_z|}\,,
\gal
where 
\ball{b9}
\cos\theta_\pm= \frac{1}{\omega}\left\{ p_z\pm \sqrt{(E-m'\Omega-\omega)^2-m'^2/R^2-M^2}\right\}\,.
\gal
We note that in order that $|\cos\theta_\pm|\le 1$, photon energy $\omega$ must not be too small.  For the states with $m'<0$ one has to take into account the requirement \eq{a15}. 

\begin{figure}[ht]
\begin{tabular}{lr}
    \includegraphics[height=5.5cm]{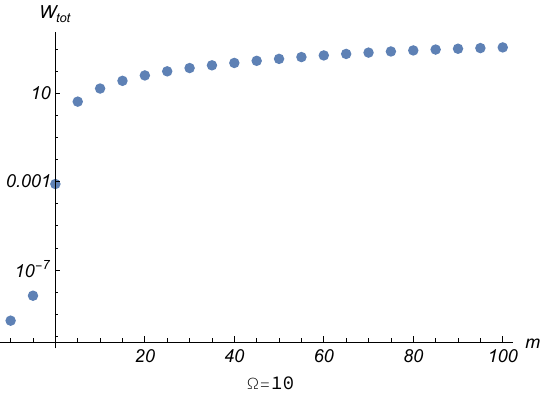} &
      \includegraphics[height=5.5cm]{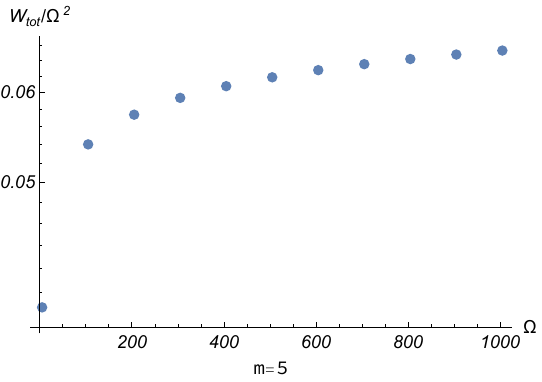}
      \end{tabular}
    \caption{Left panel: $W$ vs $m$ at $p_z=0$, right panel: $W/\Omega^2$ vs $\Omega$. In both panels $M=1$.}
    \label{fig:wtot}
\end{figure}

The radiation intensity is shown in \fig{fig:wtot} as a function of $m$ and $\Omega$. 

\section{Radiation intensity by a single particle at rest}\label{sec:rad1}

To investigate the radiation intensity analytically, consider a reference frame in which the incident particle is at rest in the axial direction: $p_z=0$, $p_z'=-\omega\cos\theta$. In particular, \eq{a15} becomes
\ball{c0}
\omega_0|\cos\theta|\le \frac{M^2}{2|m'|\Omega}\,.
\gal
In addition, we will focus on the limit $\Omega\gg M$. In this case the spectrum \eq{a2} reads
\ball{c1}
E\approx \left\{ 
\begin{array}{lc}
2m\Omega\,, & m>0\,;\\
M\,, & m=0\,;\\
\frac{M^2}{2|m|\Omega}\,, & m<0\,.
\end{array}
\right.
\qquad 
E'\approx \left\{ 
\begin{array}{lc}
\sqrt{p_z'^2+m'^2\Omega^2}+m'\Omega\,, & m'>0\,;\\
\sqrt{p_z'^2+M^2}\,, & m'=0\,;\\
\frac{p_z'^2+M^2}{2|m'|\Omega}\,, & m'<0\,.
\end{array}
\right.
\gal
In order that $E>E'$, the range of $m'$ must be $m'\le m-1$. 

We distinguish three cases: (A) $m>0$, (B) $m=0$ and (C) $m<0$. 

\subsection{$m>0$}

\begin{figure}[ht]
    \centering
    \includegraphics[height=5.5cm]{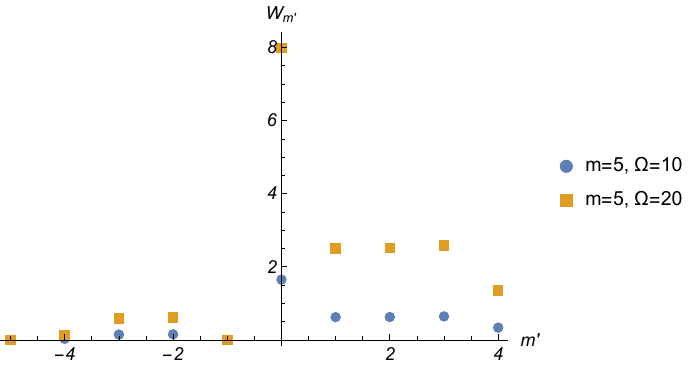}
    \caption{(Color online) The radiation intensity $W_{m'}$ of a single term in \eq{b5} for $m=5$ vs $m'$. The total intensity $W$ is the sum of all $W_{m'}$. The blue and orange symbols overlap at $m'=-1$.}
    \label{fig:WA}
\end{figure}

\fig{fig:WA} shows the dependence of a single term in the sum \eq{b5} on $m'$ for $m>0$. The mode with $m'=0$ is enhanced over those with $m'>0$ which in turn are enhanced as compared to the modes $m'<0$. 
To understand the dynamics in each case and find an appropriate approximation we split the summation over $m'$ into three parts: (A1) $m'\le -1$, (A2) $m'\ge 1$ and (A3) $m'=0$.  

\subsubsection{$m'<0$}
\begin{figure}[ht]
\begin{tabular}{lr}
    \includegraphics[height=4.5cm]{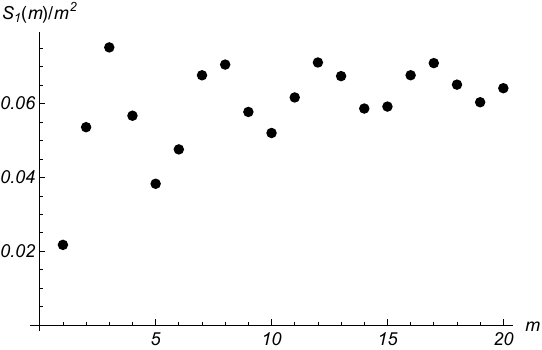} &
      \includegraphics[height=4.5cm]{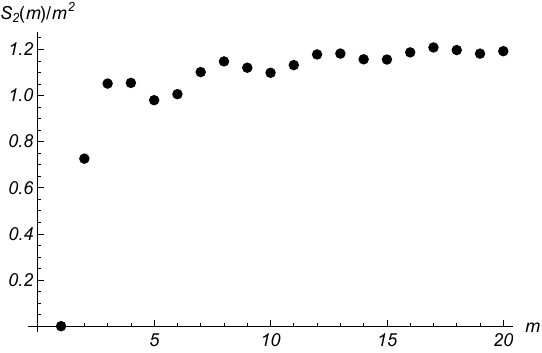}
      \end{tabular}
    \caption{Left panel: $S_1(m)/m^2$, right panel: $S_2(m)/m^2$ appearing in \eq{c9} and \eq{c14} respectively. }
    \label{fig:Wm}
\end{figure}

Using \eq{c1},\eq{b6},\eq{b7} in \eq{b5} and keeping the leading terms in $M/\Omega$ we obtain
\bal
W^{A1}= \frac{e^2}{2\pi}\sum_{m'=-\infty}^{-1}\int_0^\pi \frac{1}{16\pi  EE'}
\frac{\delta(2m\Omega-\omega)}{\frac{\omega\cos^2\theta}{E'-m'\Omega}+1}
\left[ \frac{(m+m')^2}{R^2}+\sin^2\theta\cos^2\theta\omega^2\right]\nonumber\\
\times J_{m-m'}^2(k_\bot R)\omega^2 \eta\left(M^2+2m'\Omega\omega|\cos\theta|\right)
d\omega \sin\theta d\theta \label{c3}\\
=\frac{e^2}{2\pi}\sum_{m'=-\infty}^{-1}2\int_0^1
\frac{E'-m'\Omega}{16\pi   EE'(2m\Omega\cos^2\theta+ E'-m'\Omega)}
\left[ \frac{(m+m')^2}{R^2}+(1-x^2)x^2(2m\Omega)^2\right]\nonumber\\
\times J_{m-m'}^2\left(2m\sqrt{1-x^2}\right)(2m\Omega)^2 \eta\left(M^2+2m'\Omega\omega x\right)
 dx\,,\label{c4}
\gal
where $\eta$ is the step function accounting for \eq{c0}, $x=\cos\theta$ and we took advantage of the fact that the integrals over $0\le x\le 1$ and $-1\le x \le 0$ are equal. In view of \eq{c1} and \eq{c0}, the angular integration is confined to the small region  
\ball{c5}
x< \frac{M^2}{4\Omega^2m|m'|} \ll 1\,.
\gal
Bearing this in mind and using  \eq{c1} we can then write $\omega_0\cos^2\theta+E'\approx \frac{M^2}{2|m'|\Omega}\ll \Omega$. After expanding the remaining terms in $x$ we obtain 
\bal
W^{A1}
=\frac{e^2\Omega^2}{8\pi }\sum_{j=1}^\infty (m-j)^2J^2_{m+j}(2m)= \frac{e^2\Omega^2}{8\pi }S_1(m) \,,\label{c7}
\gal
where $j= -m'$ and we defined 
\ball{c9}
S_1(m)= \sum_{j=1}^\infty (m-j)^2J^2_{m+j}(2m)\,.
\gal
This function is shown in \fig{fig:Wm}.

\subsubsection{$m'>0$}

Now consider the final states with $m'>0$. In this case the constraint \eq{c0} does not apply. It implies that we can safely neglect $M$ in \eq{b7}, assuming that the integral over $\theta$ is finite at $\theta=\pi/2$ (which is indeed the case as will be seen shortly). Thus,
\ball{c11}
\omega_0\approx \frac{\Omega}{\sin^2\theta}\left\{ 2m-m'-\sqrt{(2m-m')^2-4m(m-m')\sin^2\theta}\right\}\equiv \Omega z\,,
\gal
and $E'= 2\Omega m -\omega_0$. A simple calculation yields 
\ball{c13}
W^{A2}=\frac{e^2\Omega^2}{32\pi}S_2\left(m\right)\,,
\gal
where we defined the function 
\ball{c14}
S_2(m)=  &\sum_{m'=1}^{m-1}\int_0^\pi \frac{(2m-m'-z)\left[ (m+m')^2+z^2\sin^2\theta\cos^2\theta \right]}{m(2m-z)(2m-m'-z\sin^2\theta)}J^2_{m-m'}(z\sin\theta)z^2\sin\theta d\theta\,
\gal
which depends only on $m$. The numerical calculation of $S_2$ is shown in \fig{fig:Wm}.

\subsubsection{$m'=0$}
In this case the photon frequencies are  
\ball{c16}
\omega_0= \frac{2m\Omega}{\sin^2\theta}\left\{ 1-\sqrt{\cos^2\theta+\sin^2\theta\frac{M^2}{4m^2\Omega^2}}\right\}
\gal
and the intensity is
\ball{c17}
W^{A3}= \frac{e^2}{16\pi (2m\Omega)}
\int_0^\pi \frac{m^2\Omega^2+\omega_0^2\sin^2\theta\cos^2\theta}{\omega_0\cos^2\theta+\sqrt{\omega_0^2\cos^2\theta+M^2}}
J_m^2(\omega_0R\sin\theta)\omega_0^2\sin\theta d\theta\,.
\gal 
The integrand peaks at small $x=\cos\theta$ which gives rise to the logarithmically enhanced contribution. In this leading logarithmic approximation we write
\ball{c18} 
W^{A3}\approx  \frac{e^2}{16\pi }2
\int_0^1 \frac{m^2\Omega^2}{x^2+\sqrt{x^2+\frac{M^2}{4m^2\Omega^2}}}
J_m^2(2m) dx\approx \frac{e^2\Omega^2}{8\pi}m^2J_m^2(2m)\log \frac{4m\Omega}{M}\,.
\gal

\subsection{$m=0$}

In this case all finial states have $m'<0$. Using \eq{b5}-\eq{b9} and replacing $-m'=j$ we get
\ball{c20}
W^B
=\sum_{j=1}^{\infty}\frac{e^2(E'-m'\Omega)}{16\pi E E'(E'-m'\Omega+\omega_0\cos^2\theta)}
\left[\frac{j^2}{R^2}+\omega_0^2\sin^2\theta\cos^2\theta\right]
\eta\left(M^2-2j\Omega\omega|\cos\theta|\right)\nonumber\\
\times J^2_j(\omega_0R\sin\theta)\omega^2_0\sin\theta d\theta\,,
\gal
where, considering that $x=\cos\theta\ll 1$ (see \eq{c5}) we approximate:
\ball{c21}
\omega_0\approx M\left( 1-\frac{M}{2j\Omega }(1+x^2)\right)\,.
\gal
It follows that $E'+\omega_0\cos^2\theta+j \Omega \approx j\Omega$.
We are now left  with a trivial integral over $x$ which yields:
\ball{c25}
W^B=\frac{e^2\Omega^2}{8\pi}\sum_{j=1}^\infty j^2 J_j^2(MR)
\gal
Expanding the Bessel function and keeping only the leading term with $j=1$ we finally obtain
\ball{c27}
W^{B}= \frac{e^2 M^2}{32\pi}\,.
\gal

\subsection{$m<0$}

This is the most unusual case because the fermion transitions occur between the levels with negative $m$ and $m'$ which correspond to $0<E'<E<M$. First deal with the delta-function:
\ball{c31}
\delta\left(-\frac{M^2}{2m\Omega}+\frac{\omega^2\cos^2\theta +M^2}{2m'\Omega}-\omega\right)
=\frac{1}{\frac{\omega \cos^2\theta}{m'\Omega}-1}\delta\left(\omega-\frac{M^2(|m'|-|m|)}{2|m'||m|\Omega}\right)\,.
\gal
were we used \eq{c1}.
We get for the intensity using \eq{b5}:
\ball{c33}
W^C=\frac{e^2}{16\pi}\sum_{m'=-\infty}^{m-1}
\frac{2|m|\Omega}{M^2}
\frac{2|m'|\Omega}{\omega_0^2\cos^2\theta +M^2}
\frac{|m'|\Omega}{\omega_0\cos^2\theta+|m'|\Omega}\left[\frac{(m+m')^2}{R^2}+\omega^2_0\sin^2\theta \cos^2\theta \right]
\nonumber\\
\times\eta\left(M^2+2m'\Omega\omega|\cos\theta|\right)J^2_{m-m'}(\omega_0R \sin\theta)\omega_0^2 \sin\theta d\theta\,.
\gal
In view of \eq{c31}, the step function implies that $|\cos\theta|\le |m|/\nu$, where $\nu = |m'|-|m|$ is a positive integer. Since $\omega_0\sim M^2/\Omega$, we can neglect it in the denominators and  in the square brackets in  \eq{c33}. It also allows us to expand  the Bessel function  $J_\nu (\xi)\approx \frac{1}{\nu !}(\xi/2)^\nu$ and retain only the leading $\nu=1$ term. Since $|m|>1$ in this section,  $\cos\theta$ is not restricted at all in this approximation. The angular integration becomes trivial and yields
\ball{c36}
W^{C}= \frac{e^2 M^4}{192\pi  \Omega^2}\frac{(2|m|+1)^2}{|m|^3(|m|+1)^3}\,.
\gal

\subsection{Summary of $\Omega\gg M$ approximation.}

In summary, the radiation intensity at $\Omega\gg M$ is given by
\ball{c45}
W=\frac{e^2\Omega^2}{8\pi}
\left\{\begin{array}{lc}
S_1(m)+\frac{1}{4}S_2(m)+m^2 J_m^2(2m)\log \frac{4m\Omega}{M}\,, & m>0\,, \\
\frac{M^2}{4\Omega^2}\,, & m=0\,, \\
\frac{M^4}{24\Omega^4}\frac{(2|m|+1)^2}{|m|^3(|m|+1)^3}\,, & m<0\,.
\end{array}\right.
\gal
The transitions from $m>0$,  with the corresponding energy $E\approx 2\Omega m$, give the largest contribution. When $m$ is not very large, the leading channel is the transition to $m'=0$ ($E'\approx M\ll E$) given by \eq{c18}. This is also seen in \fig{fig:WA}. For  large $m$, we can use the well-known formula (see e.g.\ \textbf{9.3.15} in \cite{abramowitz1968handbook})
\ball{f19}  
J_m(2m)\approx \sqrt{\frac{2}{\pi \sqrt{3} m}}\cos\left[ m(\sqrt{3}-\pi/3)-\pi/4\right]\,, \quad m\gg 1\,,
 \gal 
to conclude that  the contribution from the transitions to $m'>0$ becomes dominant since $S_2(m)\sim m^2$. We verified that \eq{c45} is an accurate approximation of the exact formula at $p_z=0$ and $M\ll \Omega$ (plotted in \fig{fig:wtot}).

\section{Radiation by spinning ideal gas}

\subsection{Energy spectrum}

\begin{figure}[ht]
\begin{tabular}{lr}
    \includegraphics[height=5.5cm]{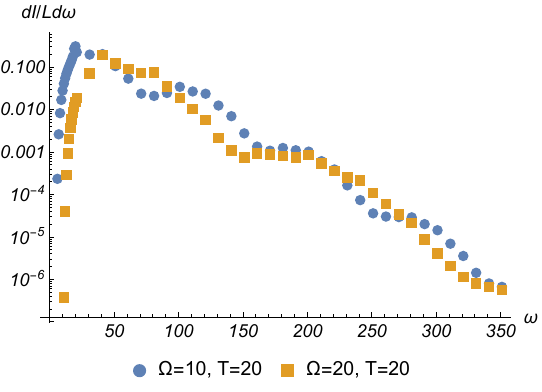} &
      \includegraphics[height=5.5cm]{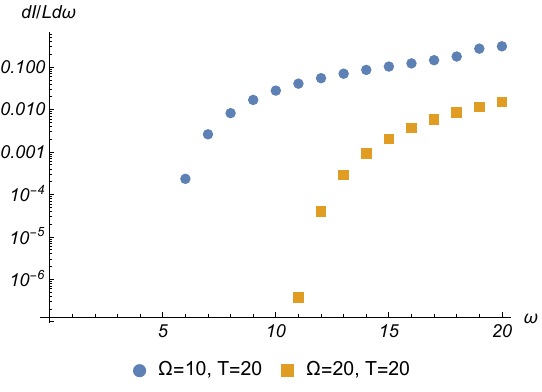}
      \end{tabular}
    \caption{(Color online) The energy spectrum of electromagnetic radiation by a thermal system. Right panel zooms into the infrared region of the left one. $M=1$. }
    \label{fig:dIdw}
\end{figure}

If an ideal Maxwell-Boltzmann gas rotates extremely rapidly, then its radiation intensity is given by 
\ball{f1}
I &= \sum_{m=-\infty}^\infty \int \frac{dp_z L}{2\pi}\int do \int_0^E d\omega \frac{dW}{dod\omega}e^{-\beta E}\,.
\gal
Using \eq{b5},\eq{b8} and \eq{b9} we obtain
\ball{f3}
\frac{dI}{Ld\omega} = \frac{e^2}{16\pi}&\sum_{m=-\infty}^\infty\sum_{\pm}\sum_{m'=-\infty}^{m-1}
\int \frac{dp_z}{2\pi}  e^{-\beta E} \frac{E'-m'\Omega}{EE'\omega |\omega \cos\theta_\pm-pz|}\nonumber\\
&\times\left[
\frac{(m+m')^2}{R^2}+\sin^2\theta_\pm(2p_z-\omega \cos\theta_\pm)^2
\right]
J_{m-m'}^2(\omega R\sin\theta_\pm)\omega^2\,.
\gal
where the integral over $p_z$ is restricted by \eq{a15} for $m<0$. The corresponding energy spectra are exhibited in \fig{fig:dIdw}. We observe that there is the threshold frequency below which photon emission is impossible. We also notice that the spectrum peaks at $\omega\sim \Omega$.

\subsection{Photon production in heavy-ion collisions}\label{sec:hic}

It is tempting to apply the model developed in this article to describe the  photon production by rotating quark-gluon plasma. We certainly realize that the plasma is not yet in the regime of extremely fast rotation $\Omega> \ell^{-1}$. On the other hand, it is already in the region of fast rotation $\Omega\sim a^{-1}$, using the notation introduced in Introduction. We therefore consider the calculation of the  photon spectrum as the no more than a back-of-the-envelop estimate and the proof of principle that rotation is a relevant effect.  

The photon emission from quark-gluon plasma, is described in terms of the following variables:  $k_T$, the photon momentum  in the plane perpendicular to the collision axis (not to be confused with $k_\bot$ defined with respect to the rotation axis), $\phi$ its azimuthal angle in that plane and $y$ its rapidity. They are related to the photon energy $\omega$ and the emission angle $\theta$ (see \cite{Tuchin:2014pka} for more details):
\ball{f9}
\omega=k_T\cosh y, \quad  \cos\theta= \sin\phi/\cosh y\,.
\gal
We now express the  photon spectrum as 
\ball{f10}
\frac{dN(k_T,y,\phi)}{k_Tdk_T d\phi dy}=  g \Delta t\frac{dI(\omega,\theta)}{\omega^2d\omega do}\,,
 \gal
where $\Delta t$ is the time interval.  In realistic case photons are emitted by fermions with two possible polarizations and three colors and three flavors which make up the degeneracy factor $g=18$.   Eq.~\eq{f10} simplifies at the midrapidity region $y=0$:
\ball{f12}
\frac{dN}{k_Tdk_T  dy}\bigg|_{y=0}=  &\frac{g \Delta t L}{k_T^2}\int_0^{2\pi}d\phi  \frac{dI}{L d\omega do}\bigg|_{\omega=k_T, \, \theta= \frac{\pi}{2}-\phi}\,.
\gal
The intensity $I$ is given by \eq{f1} and $W$ is given by \eq{b5}.  

It is now advantageous to use the delta-function in \eq{b5} to take the integral over $p_z$. To this end we write 
\ball{f14}
\delta(E-E'-\omega)= \delta\left( \sqrt{p_z^2+m^2\Omega^2+M^2}-\sqrt{(p_z-\omega\cos\theta)^2+m'^2\Omega^2+M^2}+\Delta\right)\nonumber\\
=
\frac{\delta(p_z-p_{z0})(E-m\Omega)(E-\omega-m'\Omega)}{|\omega\cos\theta (E-m\Omega)+p_z\Delta| }\,,
\gal
where we introduced a convenient notation
\ball{f16}
\Delta= (m-m')\Omega -\omega\,.
\gal
To compute $p_{z0}$ we rewrite the equation in the second delta-function in \eq{f14} as a quadratic equation for $p_z$. Of course not all its solutions necessarily satisfy the original equation. A careful examination of its two roots reveals that one root satisfies it at $\Delta>0$, while another one at $\Delta<0$. These can be combined in a single formula:
\ball{f18}
p_{z0}= &\frac{1}{2(\Delta^2-\omega^2\cos^2\theta)}\Big\{ 
-\omega\cos\theta(\omega^2\cos^2\theta -\Delta^2+m^2\Omega^2-m'^2\Omega^2)\nonumber\\
&+\Delta\sqrt{(\omega^2\cos^2\theta -\Delta^2+m^2\Omega^2-m'^2\Omega^2)^2+4(m^2\Omega^2+M^2)(\omega^2\cos^2\theta-\Delta^2)}
\Big\}\,,
\gal
provided that 
\ball{f20}
\Delta^2-\omega^2\cos^2\theta\le 0\,.
\gal
Using \eq{f16} in \eq{f20} and noting that  $|\cos\theta|\le 1$ we find that the allowed photon energies are 
\ball{f22}
\omega\ge  \frac{1}{2}(m-m')\Omega\,.
\gal 
In particular, the photon spectrum has an infrared threshold at $\omega_\text{min}= \Omega/2$. This is indeed clearly seen in \fig{fig:dIdw}. Eq.~\eq{f20} is a constraint on the allowed values of the photon emission angle: 
\ball{f24}
|\theta|\le \Theta=\arcsin \sqrt{1-\Delta^2/\omega^2}\,.
\gal
Taking all these into account we obtain the final expression for the  photon spectrum:
\ball{f26}
\frac{dN}{k_Tdk_T  dy}\bigg|_{y=0}= & \frac{g\Delta t L e^2}{4(2\pi)^3}\sum_{m=-\infty}^{\infty}\sum_{m'=-\infty}^{m-1}\int_{-\Theta}^{+\Theta} d\theta \frac{1}{EE'}e^{-E/T}\frac{(E-m\Omega)(E'-m'\Omega)}{|k_T\cos\theta (E-m\Omega)+p_z\Delta| }\nonumber\\
&\times\left[ (m+m')^2\Omega^2+\sin^2\theta(2p_{z0}-k_T\cos\theta)^2\right]J^2_{m-m'}(k_T\sin\theta\Omega^{-1})\,,
\gal
valid for $k_T=\omega$ satisfying \eq{f22}. $E$ and $E'$ are the functions of $p_{z0}$. At negative $m$ and $m'$ the angular integration is further restricted by \eq{a15}. However,  as we have seen, the negative $m$ and $m'$ give a negligible contribution. Therefore in practice the sums over $m$ and $m'$ rum only over the non-negative values. 
\begin{figure}[ht]
    \centering
    \includegraphics[height=3in]{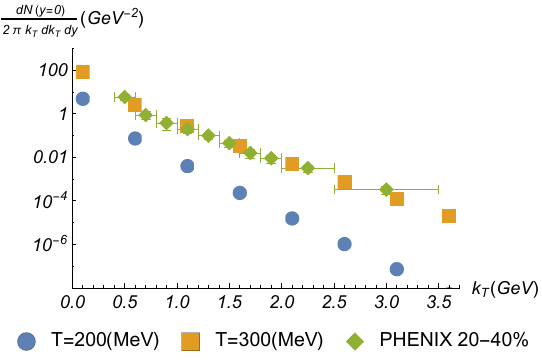}
    \caption{(Color online) Photon spectrum at two temperatures and $\Omega=0.1$~fm$^{-1}$. The data is from \cite{PHENIX:2014nkk}. $\Delta t =10$~fm/$c$, $L=10$~fm, $M=150$~MeV.}
    \label{fig:dNphoton}
\end{figure}
%

The results of the calculation are shown in \fig{fig:dNphoton} along with the experimental data on photon production in relativistic heavy-ion collisions \cite{PHENIX:2014nkk}. We observe a wondrous agreement for a particular plasma temperature. This does not of course imply that plasma rotation is the only source of photons. However, it does indicate that it must be accounted for in comprehensive phenomenological models. 
And perhaps it will help resolving the long-standing puzzle of the photon excess at low momenta. 

\section{Summary}

We posited that systems rotating with extreme angular velocity $\Omega\sim \ell^{-1}$, where $\ell$ is the mean-free-path, can be described by matter fields confined to the two dimensional cylindrical surface of radius $R= 1/\Omega$. We developed an application of this idea to the photon production by the rotating quark-gluon plasma and argued that it is consistent with the experimental observations.  

The statistical properties inside the light-cylinder are equivalent to those  of the two-dimensional ideal gas, as the interactions are screened by $R<\ell$. However, the statistical properties of the entire plasma are determined by the interaction of the light-cylinders;  the development of these ideas is left for another study.

\acknowledgments
This work  was supported in part by the U.S. Department of Energy Grants No.\ DE-SC0023692.


\bibliographystyle{apsrev4-2}
\bibliography{Biblio}

\end{document}